# Effect of Surface Termination on the Lattice Thermal Conductivity of Monolayer Ti$_3$C$_2$T$_z$ MXenes

*Hamed Gholivand[1], Shadi Fuladi[2], Zahra Hemmat[1], Amin Salehi-Khojin[1], Fatemeh Khalili-Araghi[2,*]*

[1]Department of Mechanical and Industrial Engineering, University of Illinois at Chicago, Chicago, IL, 60607, USA.

[2]Department of Physics, University of Illinois at Chicago, Chicago, IL, 60607, USA.

[*] Corresponding author: akhalili@uic.edu

**Abstract**

**Recently two-dimensional (2D) transition metal carbides and nitrides (MXenes) have gained significant attention in electronics and electrochemical energy conversion and storage devices where the heat production significantly affects the safety and performance of these devices. In this paper, we have studied the thermal transport in monolayer $Ti_3C_2T_z$, the first and most studied MXene, using density functional theory (DFT) and phonon Boltzmann transport equation and quantified the effect of surface termination (bare, fluorine and oxygen) on its lattice thermal conductivity. We found that thermal conductivity of fluorine-terminated $Ti_3C_2T_z$ (108 W/m.K) is approximately one order of magnitude higher than its oxygen-terminated counterpart (10 W/m.K). Our calculations reveal that the increased thermal conductivity for the fluorine-terminated structure is due to its enhanced specific heat and group velocity and diminished scattering rate of phonons.**

Since the discovery of the first member of two-dimensional (2D) materials, graphene[1,2], there has been an extensive effort to find new 2D materials with remarkable electronic, thermoelectric, optical, and thermal properties[1,3,4]. Besides 2D transitional metal dichalcogenides[5,6] (e.g. $MoS_2$ and $WS_2$) and hexagonal boron nitride[7], a new member of hexagonal transition-metal carbides/nitrides, called MXene has recently shown a great promise for a wide range of applications including energy storage and conversion[8–10], supercapacitors[11,12], fuel cells[13], and electronic devices[14,15]. MXene is exfoliated from the MAX phase by the etching process[16,17]. The MAX phase consists of more than 60 families of layered hexagonal transition-metal carbides/nitrides with the general formula of $M_{n+1}AX_n$ (n= 1, 2 or 3), where M is an early transition metal (Ti, Nb, V, Cr, etc.), A is mainly a group IIIA or IVA element, and X represents C and/or N. In the etching process, depending on the synthesis conditions, the A atoms are replaced by F, O, or OH[17,18] due to fluorine and $H_2O$ containing aqueous solutions; hence, the general formula of MXenes becomes $M_{n+1}X_nT_x$ where T denotes the type of surface termination composition.

Previous theoretical studies suggest that the energy storage capacities[19,20] and electronic[17] and magnetic[21] properties of MXenes strongly depend on their surface termination composition[21]. However up to now, there are only a few reports on the thermal transport properties of MXene structures such as $Ti_3C_2$[22], $Ti_2CT_2$ $(T = O, F, OH)$[23] and $T_2CO_2$ $(T = Ti, Zr, Hf, Sc)$[24]. Recently, Liu et al[25] have measured the thermal conductivity of $Ti_3C_2T_z$ as 55.88 W/m.K, where chemical composition of surface termination atoms is unknown. In this study, we have systematically investigated the effect of surface termination on the lattice thermal conductivity of monolayer $Ti_3C_2T_z$, the first and most studied MXene, using density functional theory (DFT) and iterative solution of linearized Boltzmann transport equation (BTE). Our results provide a guidance for future experimental studies and possible applications of $Ti_3C_2$ MXenes in nanoelectronic and energy storage and conversion systems[8,26].

In this study, all calculations are based on DFT using the projected augmented wave (PAW) potentials[27] method as implemented in the Vienna ab initio simulation package (VASP)[28]. The exchange-correlation functional is described by generalized gradient approximation (GGA) within the Perdew-Burke-Ernzerhof (PBE)[29] formulation. For geometry optimizations, conjugate gradient method is used while convergence threshold is set to $10^{-7}\ eV$ and $10^{-4}\ eV/Å$ for electronic and ionic relaxation, respectively. Lattice thermal properties are obtained by implementing the second- and third-order interatomic force constants (IFCs) in ShengBTE[30] code, where a q-point grid of $19 \times 19 \times 1$ is used. The second-order IFCs are calculated by PHONOPY[31] code, using a $4 \times 4 \times 1$ supercell and k-point sampling of $3\times3\times1$ for Brillouin zone where kinetic energy cutoff is set to 500 eV. The third-order IFCs are evaluated using the same supercell size and k-point meshing by considering up to six nearest-neighbor interactions. In addition, Born effective charges and dielectric tensor are calculated using VASP in order to include long-range electrostatic interactions into account.

Previous studies suggest various structural models of terminated $Ti_3C_2T_z$ based on the position of termination atoms[26]. Here, we have selected the most stable and energetically favorable configuration where termination atoms tend to sit in the hollow site of three neighboring carbon atoms[26,32]. All studied structures have the same crystal symmetry of $P\bar{3}m1$ (space group: 186). Relaxed structures are shown in Fig. 1 which are visualized using VMD[33]. Lattice constant, effective thickness (vertical distance from top-most atomic layer to the bottom-most atomic layer), and bond lengths of each structure are presented in Table 1. Calculated relaxed lattice constants and bond lengths show excellent agreement with previous theoretical studies[32,34–37]. Based on the bond lengths presented in table 1 we realize that Ti(1)-C, Ti(2)-C bond lengths decreases and increases, respectively, when terminated with either F or O. This is due to a strong localization of electrons between Ti(2) and termination atoms[34] or in other words, strong bonding that occur among the aforementioned atoms which results in the shrinkage of lattice constant *a*, and elongation of effective thickness *d*. Furthermore, unlike $Ti_3C_2O_2$, Ti(1)-C is longer than Ti(2)-C in bare and fluorine terminated $Ti_3C_2$, which is an indication of stronger covalent bonds between Ti(1)-C and Ti(2)-O in $Ti_3C_2O_2$ compared to other two structures. Finally, much shorter bond length of Ti(2)-O compared to Ti(2)-F implies that oxygen makes a stronger covalent bond with Ti(2) compared to fluorine.

Calculated phonon dispersion and projected phonon density of states (DOS) of all structures are shown in Fig. 2. Phonon dispersion relations provide important information regarding scattering rates and phonon related properties such as the phononic contributions to thermal transport[38]. Moreover, absence of imaginary frequencies in phonon dispersions indicates the dynamical stability of relaxed structures. Since there are 5 atoms in the primitive cell of bare $Ti_3C_2$, there are 15 phonon branches in the phonon dispersion relation (Fig. 2(a)), where 3 of them are acoustic and 12 are optical modes at the Γ-point. The 7 atoms of the primitive cell of fluorine and oxygen-terminated $Ti_3C_2$ give rise to 3 acoustic and 18 optical modes at the Γ-point of the Brillouin zone (Fig. 2(b), (c)). Projected phonon DOS shows a higher and more extended frequency range for oxygen in $Ti_3C_2O_2$ compared to fluorine in $Ti_3C_2F_2$, which is an indication of stronger bonding between Ti(2) and O compared to Ti(2) and F. For this reason, the bandgaps in $Ti_3C_2$ and $Ti_3C_2F_2$ dispersion curves disappear in $Ti_3C_2O_2$, resulting in higher phonon scattering rates between acoustic and high energy optical modes in $Ti_3C_2O_2$ compared to other two structures.

Figure 3(a) demonstrates the thermal conductivity of structures obtained from the iterative solution of the Boltzmann transport equation (BTE) at the temperature range of 300-600 K. In order to make a good comparison of all three structures, an 8 Å out of plane lattice constant is used for all the structures when calculating thermal conductivities. The thermal conductivity values for $Ti_3C_2$, $Ti_3C_2F_2$, and $Ti_3C_2O_2$ at room temperature are found to be 58.3 W/m.K, 107.8 W/m.K, and 10.5 W/m.K, respectively. Thermal conductivity of structures decreases by increasing the temperature and reaches to 31.2 W/m.K, 55.7 W/m.K, and 5.1 W/m.K for $Ti_3C_2$, $Ti_3C_2F_2$, and $Ti_3C_2O_2$ at 600 K, respectively. These results reveal that the thermal conductivity of fluorine-terminated $Ti_3C_2$ is an order of magnitude higher than its oxygen-terminated counterpart.

We also calculated the cumulative thermal conductivity of these structures with respect to the phonon mean free path. This is crucially important as the thermal conductivity of nanostructures can be substantially smaller than their bulk value when the characteristic lengths are comparable to the phonon mean free paths (MFPs). In order to explore the size effect in these structures, we obtained the distribution of phonon MFPs. Our results shown in Fig. 3(b) indicate that the cumulative thermal conductivity increases as the characteristic length increases until it reaches the thermodynamic limit at length $L_{diff}$, which represents the longest mean free path of the phonons[39,40]. $L_{diff}$ for $Ti_3C_2$, $Ti_3C_2F_2$, and $Ti_3C_2O_2$ is 475nm, 640nm, and 105nm, respectively. Our results also reveal that in $Ti_3C_2$, $Ti_3C_2F_2$, or $Ti_3C_2O_2$, 30% of the heat is being carried by the phonons with a mean free path greater than 170nm, 190nm, and 18.5nm, respectively.

To better understand how different surface termination compositions result in such a large variation in the thermal conductivity of $Ti_3C_2T_z$ MXenes, we calculated three effective parameters that predominantly contribute to the thermal conductivity: specific heat, group velocity and scattering rate of phonon modes. Figure S1 shows the calculated volumetric specific heat of all three structures within the temperature range of 300-700 K. Volumetric specific heat of $Ti_3C_2F_2$ and $Ti_3C_2O_2$ at room temperature are found to be 28% and 22% higher than bare $Ti_3C_2$, respectively. Group velocity of phonons which is the spatial derivative of frequency with respect to wavevector is computed and is shown in Fig. S2. Our results indicate that in $Ti_3C_2F_2$, the average group velocity of acoustic phonon modes, the main heat carriers in semiconductors and insulators[41] such as graphene[42] and h-BN[42,43], is 18% and 45% higher than bare $Ti_3C_2$ and $Ti_3C_2O_2$, respectively. Furthermore, we obtained three-phonon scattering rates of q-points for all structures, shown in Fig. 4. We found that in $Ti_3C_2O_2$ the average scattering rate of acoustic modes with frequencies up to ~240 cm$^{-1}$ is an order of magnitude higher than other structures. Due to these strong scattering rates, phonon relaxation time and consequently the mean free path of phonons in $Ti_3C_2O_2$ are much shorter than other two structures. This means that in $Ti_3C_2O_2$ structure, phonons are easily scattered and do not effectively contribute to the thermal transport. Thus, based on our results, since the specific heat of $Ti_3C_2O_2$ and $Ti_3C_2F_2$ is in the same range, lower conductivity of $Ti_3C_2O_2$ is attributed to its lower average group velocity of phonons and higher scattering rates compared to $Ti_3C_2F_2$. On the other hand, higher thermal conductivity of $Ti_3C_2F_2$ compared to bare $Ti_3C_2$ is attributed to its higher specific heat and lower scattering rate of low frequency modes.

In summary, we have systematically investigated the effect of surface termination on the lattice thermal conductivity of monolayer $Ti_3C_2T_z$ MXenes. Lattice constants and bond lengths suggest stronger bonding and interaction between oxygen and near-surface titanium atoms compared to fluorine. This strong interaction is consistent with phonon dispersion relations and density of states. It is shown that the thermal conductivity of fluorine-terminated $Ti_3C_2$ is an order of magnitude higher than its oxygen-terminated counterpart due to longer phonon relaxation time and higher group velocity of phonons. Finally, the results of cumulative thermal conductivity with respect to phonon MFPs shows that MFP of phonons contributing to the top 30% of total thermal conductivity in $Ti_3C_2F_2$ are one order of magnitude longer than that of $Ti_3C_2O_2$.

The work of F.K.-A., A.S.-K., H.G., S.F., Z.H. was supported by the National Science Foundation EFRI 2-DARE Grant 1542864. The authors acknowledge the Advanced Cyberinfrastructure for Education and Research (ACER) group at The University of Illinois at Chicago for providing HPC resources that have contributed to the research results reported within this paper. URL: https://acer.uic.edu.

**Table 1.** Optimized structures of bare Ti$_3$C$_2$, Ti$_3$C$_2$F$_2$, and Ti$_3$C$_2$O$_2$

| | lattice constant $a$ (Å) | Effective thickness $d$ (Å) | Bond lengths (Å) | | | |
|---|---|---|---|---|---|---|
| | | | Ti(1)-C | Ti(2)-C | Ti(1)- Ti(2) | Ti(2)-O/F |
| Ti$_3$C$_2$ | 3.104 | 4.641 | 2.212 | 2.064 | 2.932 | |
| Ti$_3$C$_2$F$_2$ | 3.073 | 7.213 | 2.185 | 2.081 | 2.954 | 2.167 |
| Ti$_3$C$_2$O$_2$ | 3.042 | 6.941 | 2.152 | 2.258 | 3.198 | 1.923 |

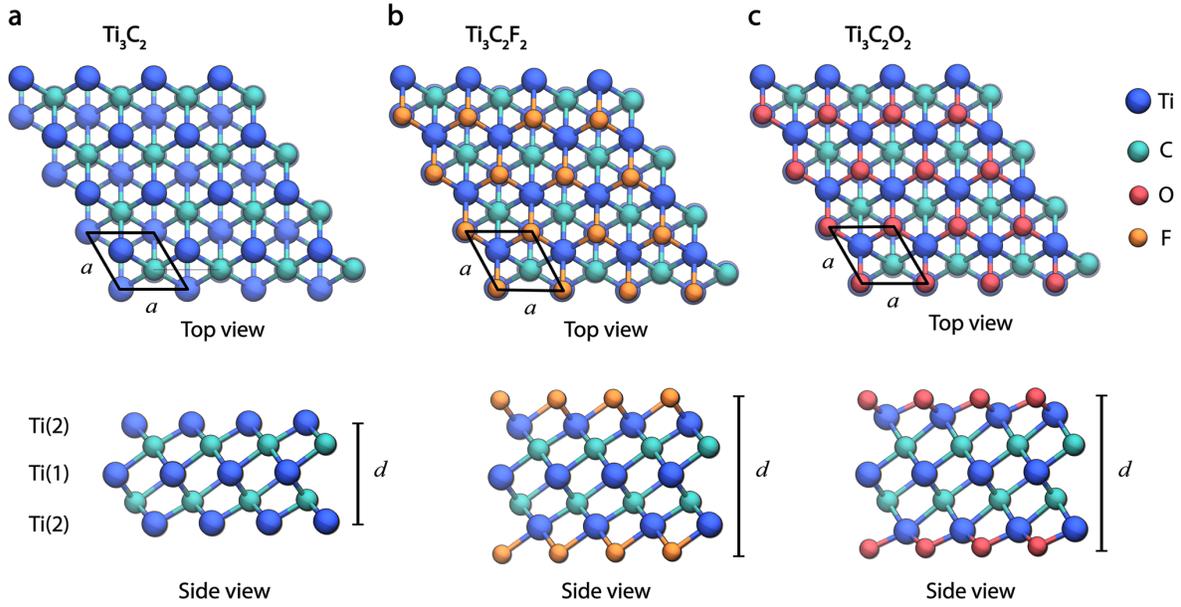

**Figure 1.** Top and side view of relaxed structures of (a) bare Ti$_3$C$_2$, (b) Ti$_3$C$_2$F$_2$, and (c) Ti$_3$C$_2$O$_2$.

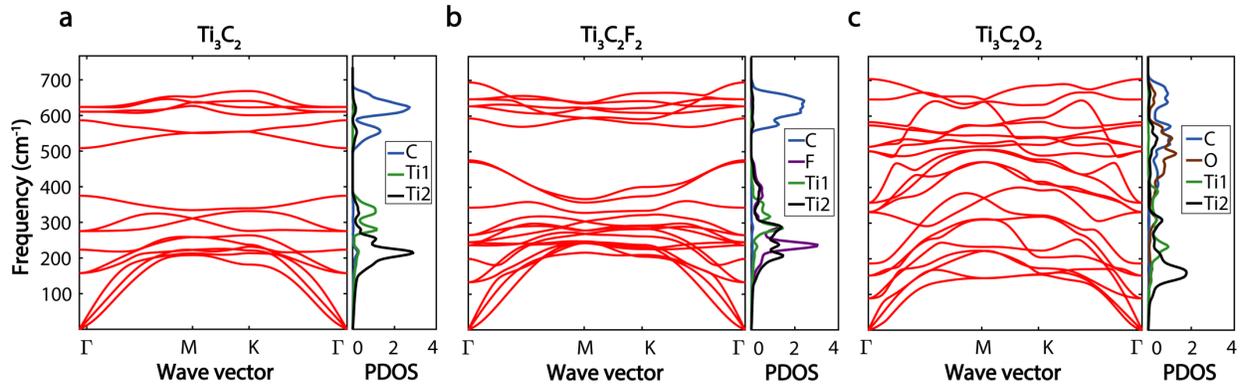

**Figure 2.** Calculated phonon dispersion relation along the high symmetry directions and projected phonon DOS of (a) Ti$_3$C$_2$, (b) Ti$_3$C$_2$F$_2$, and (c) Ti$_3$C$_2$O$_2$.

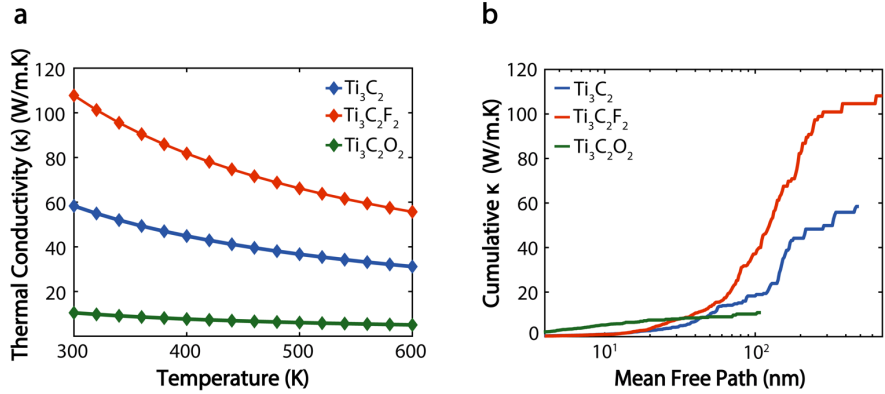

**Figure 3.** (a) Calculated thermal conductivity of bare, fluorine and oxygen-terminated $Ti_3C_2$ monolayer Mxenes, (b) Cumulative thermal conductivity of $Ti_3C_2$ monolayer Mxenes with respect to mean free paths of phonons at T=300 K.

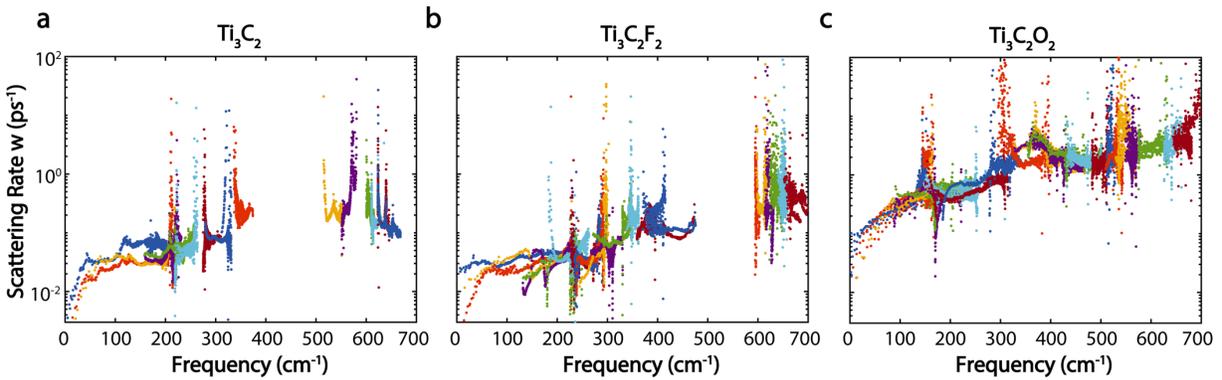

**Figure 4.** Calculated phonon scattering rates of (a) $Ti_3C_2$, (b) $Ti_3C_2F_2$, and (c) $Ti_3C_2O_2$ inside Brillouin zone using a 80×80×1 q-point grid.

# SUPPORTING INFORMATION

**Effect of Surface Termination on the Lattice Thermal Conductivity of Monolayer $Ti_3C_2T_z$ MXenes**

*Hamed Gholivand[1], Shadi Fuladi[2], Zahra Hemmat[1], Amin Salehi-Khojin[1], Fatemeh Khalili-Araghi[2,*]*

[1]Department of Mechanical and Industrial Engineering, University of Illinois at Chicago, Chicago, IL, 60607, USA.

[2]Department of Physics, University of Illinois at Chicago, Chicago, IL, 60607, USA.

[*] Corresponding author: akhalili@uic.edu

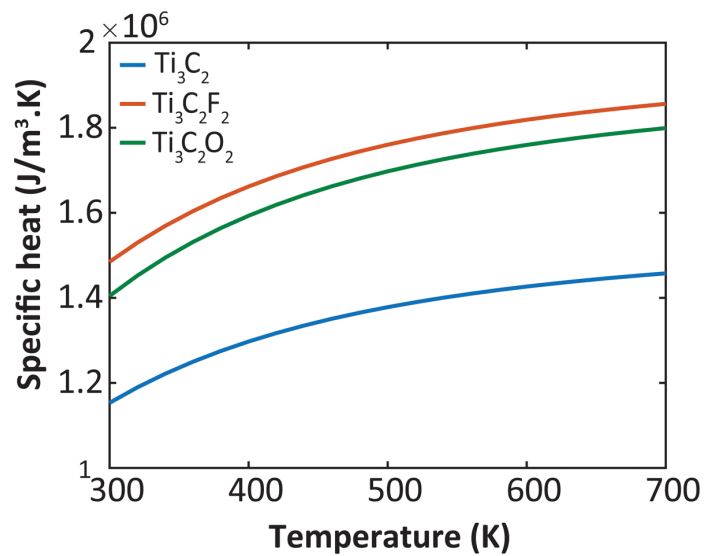

**Figure S1.** Calculated volumetric specific heat

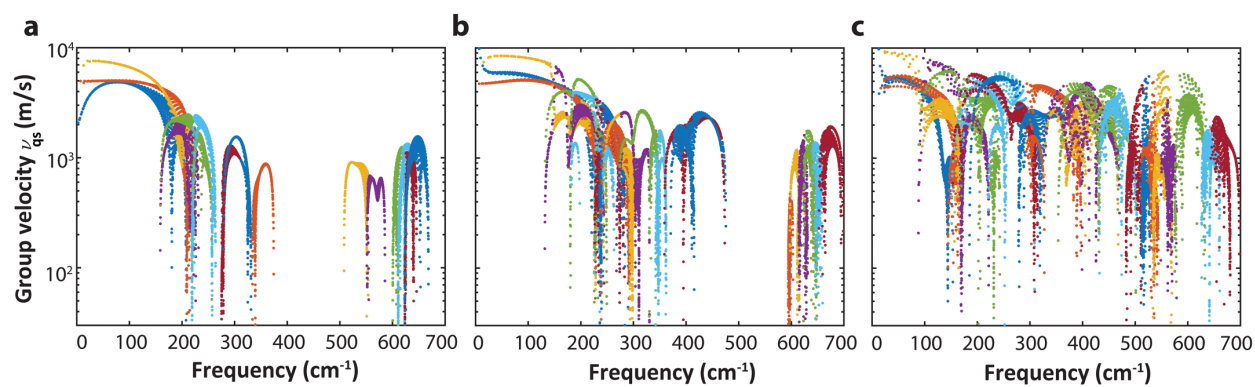

**Figure S2.** Calculated phonon group velocity of (a) $Ti_3C_2$, (b) $Ti_3C_2F_2$, and (c) $Ti_3C_2O_2$ inside Brillouin zone using a 80×80×1 q-point grid